\documentclass{amsart}
\usepackage{amsmath}

\newtheorem{theorem}{Theorem}[section]
\newtheorem{lemma}[theorem]{Lemma}
\newtheorem{proposition}[theorem]{Proposition}
\newtheorem{corollary}[theorem]{Corollary}

\theoremstyle{definition}

\theoremstyle{remark}
\newtheorem{remark}[theorem]{Remark}

\numberwithin{equation}{section}

\newcommand{\limfunc}[1]{\mathrm{#1}}
\newcommand{\tbigoplus}{\bigoplus}

\begin{document}

\title[Covariant POVM]{Positive operator valued measures covariant with respect
to an Abelian group}

\author{G.~Cassinelli}
\address{G.~Cassinelli, Dipartimento di Fisica, Universit\`a di Genova, and
I.N.F.N., Sezione  di Genova, 
 Via Dodecaneso~33, 16146 Genova, Italy.} 
\email{cassinelli@ge.infn.it}
\author{E.~De Vito}
\address{E.~De Vito, Dipartimento di Matematica, Universit\`a di Modena e
Reggio Emilia, Via Campi 213/B, 41100 Modena, Italy, and
I.N.F.N., Sezione di Genova, Via
Dodecaneso~33, 16146 Genova, Italy.}
\email{devito@unimo.it}
\author{A.~Toigo}
\address{A.~Toigo, Dipartimento di Fisica, Universit\`a di Genova, and
I.N.F.N., Sezione di Genova, Via Dodecaneso~33, 16146 Genova, Italy.}
\email{toigo@ge.infn.it}

\date{27 December 2002}
 
 
 \begin{abstract}
Given a unitary representation $U$ of an Abelian group $G$ and a subgroup $H$, we characterise the positive operator valued measures based on the
quotient group $G/H$ and covariant with respect to $U$. 
\end{abstract}

\maketitle

\setlength\arraycolsep{2pt}

\section{Introduction}

\noindent Let $G$ be a topological group and $H$ a closed subgroup. Given a
unitary representation $U$ of $G$, it is of interest in Quantum Mechanics
and in wavelet analysis to describe the positive operator valued measures $ 
Q $ defined on the quotient space $G/H$ and covariant with respect to $U$.
The generalised imprimitivity theorem  assures that $Q$ exists if
 and only if there exists a unitary
representation $\sigma $ of $H$ such that 
 $U$ is a subrepresentation of the
representation unitarily induced by $\sigma $.
 
In his seminal papers~\cite{holevo1} and~\cite{holevo2}, Holevo classifies
the covariant
 positive operator valued measures if $G$ is of type I and
$H=\{e\}$, and if $G$
 is compact and $H$ is arbitrary.

In this paper we extend the above result to the case $G$ Abelian and $H$
arbitrary. We describe the covariant positive operator valued measures in
terms of a family $W_{x}:E_{x}\rightarrow E$ of isometries, where the index $ 
x$ runs over the dual group of $G$, $\dim E_{x}$ equals the multiplicity of
the character $x$ in $U$ and $E$ is a fixed (infinite dimensional) Hilbert
space.
As a byproduct, we define a unitary operator $\Sigma$ that diagonalises
the representation of $G$ unitarily induced by a representation of $H$ with
uniform multiplicity.
  
As an application of our characterisation, in the final section we give
three examples.

\begin{enumerate}
\item  The \emph{regular} representation of the real line, where the
positive operator valued measure describes the \emph{position observable} in
one dimension.

\item  The \emph{number}-representation of the torus, where the positive
operator valued measure describes the \emph{phase observable}.

\item  The tensor product of two \emph{number}-representations of the torus,
where the positive operator valued measure describes the \emph{phase
difference observable}.
\end{enumerate}

\section{Notations}

In this paper, by \emph{Hilbert space} we mean a separable complex Hilbert
space with scalar product $\left\langle \cdot ,\cdot \right\rangle $ linear
in the first argument, by \emph{group} we mean a locally compact second
countable Abelian group and by \emph{representation} a continuous unitary
representation of a group acting on a Hilbert space. If $X$ is a locally
compact second countable topological space, we denote by $\mathcal{B}\left(
X\right) $ the Borel $\sigma $-algebra of $X$ and by $C_{c}\left( X\right) $
the space of continuous complex functions on $X$ with compact support. By 
\emph{measure} we mean a positive measure defined on $\mathcal{B}\left(
X\right) $ and finite on compact sets.

In the sequel we shall use rather freely basic results of Harmonic Analysis
on Abelian groups, as exposed, for example, in Refs.~\cite{Dieu2}
and~\cite{Foll}.
 
We fix a group $G$ and a closed subgroup $H$. We denote by $\widehat{G}$ and 
$\widehat{H}$ the corresponding dual groups and by $\left\langle
x,g\right\rangle$ the canonical pairing.

We denote by 
\begin{equation*}
q:G\longrightarrow G/H\text{,}\qquad q\left( g\right) =\dot{g}
\end{equation*}
the canonical projection onto the quotient group $G/H$. If $a\in G$ and $\dot{g}\in G/H$, we let $a\left[ 
\dot{g}\right]=q\left( ag\right) =\dot{a}\dot{g}$ be the natural action of $ 
a $ on the point $\dot{g}$.

Let $H^{\perp }$ be the annihilator of $H$ in $\widehat{G}$, that is 
\begin{equation*}
H^{\perp }=\left\{ y\in \widehat{G}\mid \left\langle y,h\right\rangle =1 
\text{ }\forall h\in H\right\} \text{.}
\end{equation*}
The group $H^{\perp }$ is a closed subgroup of $\widehat{G}$ and $
\widehat{G/H}$ can be identified (and we will do) with $H^{\perp }$ by means
of 
\begin{equation*}
\left\langle y,\dot{g}\right\rangle :=\left\langle y,g\right\rangle \qquad
\forall y\in H^{\perp }\text{, }\forall \dot{g}\in G/H\text{.}
\end{equation*}

Since $H^{\perp }$ is closed, we can consider the quotient group $\widehat{G} 
/H^{\perp }$. We denote by 
\begin{equation*}
\pi :\widehat{G}\longrightarrow \widehat{G}/H^{\perp }\text{,}\qquad \pi
\left( x\right) =\dot{x}
\end{equation*}
the canonical projection. The group $\widehat{H}$ can be identified (and we
will do) with the quotient group $\widehat{G}/H^{\perp }$ by means of 
\begin{equation*}
\left\langle \dot{x},h\right\rangle :=\left\langle x,h\right\rangle \qquad
\forall \dot{x}\in \widehat{G}/H^{\perp }\text{, }\forall h\in H\text{.}
\end{equation*}

Let $\mu _{G}$, $\mu _{H}$ and $\mu _{G/H}$ be fixed Haar measures on $G$, $ 
H $ and $G/H$, respectively.

We denote by $\mu _{H^{\perp }}$\ the Haar measure on $H^{\perp }$ such that
the Fourier-Plancherel cotransform $\overline{\mathcal{F}}_{G/H}$ is unitary
from $L^{2}\left( G/H,\mu _{G/H}\right) $ onto $L^{2}\left( H^{\perp },\mu
_{H^{\perp }}\right) $, where $\overline{\mathcal{F}}_{G/H}$ is given by 
\begin{equation*}
\left( \overline{\mathcal{F}}_{G/H}f\right) \left( y\right)
=\int_{G/H}\left\langle y,\dot{g}\right\rangle f\left( \dot{g}\right) \text{d}\mu _{G/H}\left( \dot{g}\right) 
\qquad y\in H^{\perp },
\end{equation*}
for all $f\in \left( L^{1}\cap L^{2}\right)
\left( G/H,\mu _{G/H}\right) $. 

Given $\varphi \in C_{c}\left( \widehat{G}\right) $, let 
\begin{equation*}
\widetilde{\varphi} \left( \dot{x}\right) :=\int_{H^{\perp }}\varphi \left(
xy\right) \text{d}\mu _{H^{\perp }}\left( y\right) \qquad \forall \dot{x}\in 
\widehat{G}/H^{\perp }\text{.}
\end{equation*}
It is well known that $\widetilde{\varphi}$ is in $C_{c}\left( \widehat{G} 
/H^{\perp}\right) $ and that $\widetilde{\varphi}\geq 0$ if $\varphi \geq 0 $. Given a measure $\nu $ on $\widehat{G}/H^{\perp}$, the map 
\begin{equation}
C_{c}\left( \widehat{G}\right) \ni \varphi \longmapsto \int_{\widehat{G} 
/H^{\perp }}\widetilde{\varphi} \left( \dot{x}\right) \text{d}\nu \left( 
\dot{x}\right) \in \mathbb{R}  \label{def. nu tilde}
\end{equation}
is linear and positive. Hence, by Riesz-Markov theorem, there is a unique
measure $\widetilde{\nu }$ on $\widehat{G}$ such that 
\begin{equation*}
\int_{\widehat{G}}\phi \left( x\right) \text{d}\widetilde{\nu }\left(
x\right) =\int_{\widehat{G}/H^{\perp }}\text{d}\nu \left( \dot{x}\right)
\int_{H^{\perp }}\phi \left( xy\right) \text{d}\mu _{H^{\perp }}\left(
y\right)
\end{equation*}
for all $\phi \in L^{1}\left( \widehat{G},\widetilde{\nu}\right) $. One can
check that the correspondence $\nu\longmapsto\widetilde{\nu}$ preserves
equivalence and orthogonality of measures.

Given a finite measure $\mu $ on $\widehat{G}$, we denote by $\mu ^{\pi }$
the image measure of $\mu$ with respect to $\pi$, i.e.~the measure on $ 
\widehat{G}/H^{\perp }$ given by 
\begin{equation*}
\mu ^{\pi }\left( A\right) =\mu \left( \pi ^{-1}\left( A\right) \right)
\qquad \forall A\in \mathcal{B}\left( \widehat{G}/H^{\perp }\right) \text{.}
\end{equation*}

We fix a representation $U$ of $G$ acting on a Hilbert space $\mathcal{H}$.
Let $Q$ be a positive operator valued measure (POVM) defined on $G/H$ and acting on $\mathcal{H}$.
If $Q$ satisfies the following properties
\begin{enumerate}
\item  \label{cond. 1}$Q\left( G/H\right) =I$,

\item  \label{cond. 2}for all $X\in \mathcal{B}\left( G/H\right) $, 
\begin{equation*}
U\left( g\right) Q\left( X\right) U\left( g^{-1}\right) =Q\left( g\left[ X 
\right] \right) \qquad \forall g\in G,
\end{equation*}
\end{enumerate}
it is called \emph{covariant} and $(U,Q)$ is said to be a \emph{covariance system}. 
In particular, if $Q$ is a projective measure, $\left(
U,Q\right) $ is called an \emph{imprimitivity system}.

For $\omega \in C_{c}\left( G/H\right) $, we define the operator 
\begin{equation*}
M\left( \omega \right) :=\int_{G/H}\omega \left( \dot{g}\right) \text{d} 
Q\left( \dot{g}\right) \text{.}
\end{equation*}
The map $\omega\longmapsto M\left( \omega \right)$ defines uniquely the POVM 
$Q$. In the following we use $M$ instead of $Q$.

Finally, given a representation $\sigma $ of $H$, we denote by $\left( 
\limfunc{ind}_{H}^{G}\left( \sigma \right) ,M_{0}\right) $ the imprimitivity
system induced by $\sigma $ from $H$ to $G$.

The aim of this paper is to describe all the positive operator valued
measures covariant with respect to $U$. The generalised imprimitivity
theorem (see, for example, Refs.~\cite{Cass} and~\cite{Catt}) states that

\begin{theorem}
A POVM $M$ based on $G/H$ and acting on $\mathcal{H}$ is covariant with
respect to $U$ if and only if there exists a representation $\sigma $ of $H$
and an isometry $W$ intertwining $U$ with $\limfunc{ind}_{H}^{G}\left(
\sigma \right) $ such that 
\begin{equation*}
M\left( \omega \right) =W^{\ast }{M_{0}}\left( \omega \right) W
\end{equation*}
for all $\omega \in C_{c}\left( G/H\right) $.
\end{theorem}

If $\sigma ^{\prime }$ is another representation of $H$ such that $\sigma $
is contained (as subrepresentation) in $\sigma $, then $\left( \limfunc{ind} 
_{H}^{G}\left( \sigma \right) ,M_{0}\right) $ is contained (as imprimitivity
system) in $\left( \limfunc{ind}_{H}^{G}\left( \sigma ^{\prime }\right) ,M^{\prime }_{0}\right) $. Hence, we can always assume that $\sigma $ in the
previous theorem has infinite multiplicity.

Moreover, there exist a measure $\nu $ on $\widehat{G}/H^{\perp }$ and an
infinite dimensional Hilbert space $E$ such that, up to a unitary
equivalence, $\sigma $ acts diagonally on $L^{2}\left( \widehat{G}/H^{\perp
},\nu ;E\right) $. The first step of our construction is to diagonalise the
representation $\limfunc{ind}_{H}^{G}\left( \sigma \right) $.

\section{Diagonalisation of $\limfunc{ind}_{H}^{G}\left( \sigma \right)  $}

In this section, given a representation of $H$ with uniform multiplicity, we
diagonalise the corresponding induced representation.

Let $\nu $ be a measure on $\widehat{G}/H^{\perp }$ and $E$ be a Hilbert
space. Let $\sigma ^{\nu }$ be the diagonal representation of $H$ acting on
the space $L^{2}\left( \widehat{G}/H^{\perp },\nu ;E\right) $, that is

\begin{equation*}
\left( \sigma ^{\nu }\left( h\right) \xi \right) \left( \dot{x}\right)
=\left\langle \dot{x},h\right\rangle \xi \left( \dot{x}\right) \qquad \dot{x} 
\in \widehat{G}/H^{\perp }{\text{,}}
\end{equation*}
where $h\in H$.

We denote by $\mathcal{H}^{\nu }$ the space of functions $f:G\times \widehat{ 
G}/H^{\perp }\longrightarrow E$ such that

\begin{itemize}
\item  $f$ is weakly $\left( \mu _{G}\otimes \nu \right) $-measurable;

\item  for all $h\in H$, 
\begin{equation}
f\left( gh,\dot{x}\right) =\overline{\left\langle \dot{x},h\right\rangle } 
f\left( g,\dot{x}\right) \qquad \forall \left( g,\dot{x}\right) \in G\times 
\widehat{G}/H^{\perp }\text{;}  \label{inv}
\end{equation}

\item  
\begin{equation*}
\int_{G/H\times \widehat{G}/H^{\perp }}\left\| f\left( g,\dot{x}\right)
\right\| _{E}^{2}\,\text{d}\left( \mu _{G/H}\otimes \nu \right) \left( 
\dot{g},\dot{x}\right) <+\infty \text{.}
\end{equation*}
\end{itemize}

\noindent We identify functions in $\mathcal{H}^{\nu }$ that are equal $ 
\left( \mu _{G}\otimes \nu \right) $-a.e.. Let $G$ act on $\mathcal{H}^{\nu
} $ as 
\begin{equation*}
\left( \lambda ^{\nu }\left( a\right) f\right) \left( g,\dot{x}\right)
:=f\left( a^{-1}g,\dot{x}\right) \qquad \left( g,\dot{x}\right) \in G\times 
\widehat{G}/H^{\perp }
\end{equation*}
for all $a\in G$. Define 
\begin{equation*}
\left( M_{0}^{\nu }\left( \omega \right) f\right) \left( g,\dot{x}\right)
:=\omega \left( \dot{g}\right) f\left( g,\dot{x}\right) \qquad \left( g,\dot{ 
x}\right) \in G\times \widehat{G}/H^{\perp }
\end{equation*}
for all $f \in \mathcal{H}^{\nu } $, $\omega \in C_{c}\left( G/H\right) $.

One can easily prove the following fact.

\begin{proposition}
The space $\mathcal{H}^{\nu }$ is a Hilbert space with respect to the inner
product 
\begin{equation*}
\left\langle f_{1},f_{2}\right\rangle _{\mathcal{H}^{\nu }}=\int_{G/H\times 
\widehat{G}/H^{\perp }}\left\langle f_{1}\left( g,\dot{x}\right)
,f_{2}\left( g,\dot{x}\right) \right\rangle _{E}\text{d}\left( \mu
_{G/H}\otimes \nu \right) \left( \dot{g},\dot{x}\right) \text{.}
\end{equation*}
If $\varphi \in C_{c}\left( G\times \widehat{G}/H^{\perp };E\right) $, let 
\begin{equation*}
f_{\varphi }\left( g,\dot{x}\right) :=\int_{H}\left\langle \dot{x} 
,h\right\rangle \varphi \left( gh,\dot{x}\right) \text{d}\mu _{H}\left(
h\right) \qquad \forall \left( g,\dot{x}\right) \in G\times \widehat{G} 
/H^{\perp }\text{.}
\end{equation*}
Then $f_{\varphi }$ is a continuous function in $\mathcal{H}^{\nu }$ such
that $\left( q\times \limfunc{id}_{\widehat{G}/H^{\perp }}\right) \left( 
\limfunc{supp}f_{\varphi }\right) $ is compact, and the set 
\begin{equation*}
\mathcal{H}_{0}^{\nu }=\left\{ f_{\varphi }\mid \varphi \in C_{c}\left(
G\times \widehat{G}/H^{\perp };E\right) \right\}
\end{equation*}
is a dense subspace of $\mathcal{H}^{\nu }$. The couple $\left( \lambda
^{\nu },M_{0}^{\nu }\right) $ is the imprimitivity system induced by $\sigma
^{\nu }$ from $H$ to $G$.
\end{proposition}

We now diagonalise the representation $\lambda ^{\nu }$. First of all, we
let $\widetilde{\nu }$ be the measure defined in $\widehat{G}$ by Eq.~(\ref{def.
nu tilde}). Let $\Lambda ^{\nu }$ be the diagonal representation of $G$
acting on $L^{2}\left( \widehat{G},\widetilde{\nu };E\right) $ as 
\begin{equation*}
\left( \Lambda ^{\nu }\left( g\right) \phi \right) \left( x\right)
=\left\langle x,g\right\rangle \phi \left( x\right) \qquad x\in \widehat{G}
\end{equation*}
for all $g\in G$.

Moreover, given $\phi :\widehat{G}\longrightarrow E$ and fixed $x\in 
\widehat{G}$, define $\phi _{x}$ from $H^{\perp }$ to $E$ as 
\begin{equation*}
\phi _{x}\left( y\right) :=\phi \left( xy\right) \qquad \forall y\in
H^{\perp }\text{.}
\end{equation*}

\begin{theorem}
There is a unique unitary operator $\Sigma $ from $\mathcal{H}^{\nu }$ onto $ 
L^{2}\left( \widehat{G},\widetilde{\nu };E\right) $ such that, for all $f\in 
\mathcal{H}_{0}^{\nu }$, 
\begin{equation}
\left( \Sigma f\right) \left( x\right) =\int_{G/H}\left\langle
x,g\right\rangle f\left( g,\dot{x}\right) \text{\textnormal{d}}\mu
_{G/H}\left( \dot{g}\right) \qquad x\in \widehat{G}\text{.}
\label{Formula di sigma}
\end{equation}
The operator $\Sigma $ intertwines $\lambda ^{\nu }$ with $ 
\Lambda ^{\nu }$. Moreover, 
\begin{equation}
\left( \Sigma ^{\ast }\varphi \right) \left( g,\dot{x}\right)
=\int_{H^{\perp }}\overline{\left\langle xy,g\right\rangle }\varphi \left(
xy\right) \text{\textnormal{d}}\mu _{H^{\perp }}\left( y\right) \qquad
\left( g,\dot{x}\right) \in G\times \widehat{H}  \label{star}
\end{equation}
for all $\varphi \in C_{c}\left( \widehat{G};E\right) $.
\end{theorem}

\begin{proof}
We first define $\Sigma $ on $\mathcal{H}_{0}^{\nu }$. Let $f\in \mathcal{H} 
_{0}^{\nu }$. Fixed $x\in \widehat{G}$, by virtue of Eq.~(\ref{inv}) the
function 
\begin{equation*}
g\longmapsto \left\langle x,g\right\rangle f\left( g,\dot{x}\right) 
\end{equation*}
depends only on the equivalence class $\dot{g}$ of $g$ and we let $f^{x}$ be
the corresponding map on $G/H$. Due to the properties of $f$, $f^{x}$ is
continuous and has compact support, so it is $\mu _{G/H}$-integrable and we
define $\Sigma f$ by means of Eq.~(\ref{Formula di sigma}).

We claim that $\Sigma f$ is in $L^{2}\left( \widehat{G},\widetilde{\nu } 
;E\right) $ and $\left\| \Sigma f\right\| _{L^{2}\left( \widehat{G}, 
\widetilde{\nu };E\right) }=\left\| f\right\| _{\mathcal{H}^{\nu }}$. Since
the map 
\begin{equation*}
\left( x,\dot{g}\right) \longmapsto f^{x}\left( \dot{g}\right) 
\end{equation*}
is continuous, by a standard argument $\Sigma f$ is continuous. Moreover, if 
$x\in \widehat{G}$ and $y\in H^{\perp }$, 
\begin{eqnarray*}
\left( \Sigma f\right) \left( xy\right)  &=&\int_{G/H}\left\langle
xy,g\right\rangle f\left( g,\dot{x}\right) \text{\textnormal{d}}\mu
_{G/H}\left( \dot{g}\right)  \\
&=&\int_{G/H}\left\langle y,\dot{g}\right\rangle \left\langle
x,g\right\rangle f\left( g,\dot{x}\right) \text{\textnormal{d}}\mu
_{G/H}\left( \dot{g}\right)  \\
&=&\overline{\mathcal{F}}_{G/H}\left( f^{x}\right) \left( y\right) \text{.}
\end{eqnarray*}
Indeed, 
\begin{eqnarray*}
\left\| \Sigma f\right\| _{L^{2}\left( \widehat{G},\widetilde{\nu };E\right)
}^{2} &=&\int_{\widehat{G}}\left\| \Sigma f\left( x\right) \right\| _{E}^{2} 
\text{d}\widetilde{\nu }\left( x\right)  \\
&=&\int_{\widehat{G}/H^{\perp }}\text{d}\nu \left( \dot{x}\right)
\int_{H^{\perp }}\left\| \left( \Sigma f\right) \left( xy\right) \right\|
_{E}^{2}\text{d}\mu _{H^{\perp }}\left( y\right)  \\
&=&\int_{\widehat{G}/H^{\perp }}\text{d}\nu \left( \dot{x}\right)
\int_{H^{\perp }}\left\| \overline{\mathcal{F}}_{G/H}\left( f^{x}\right)
\left( y\right) \right\| _{E}^{2}\text{d}\mu _{H^{\perp }}\left( y\right)  \\
&&(\text{unitarity~of~}\overline{\mathcal{F}}_{G/H}) \\
&=&\int_{\widehat{G}/H^{\perp }}\text{d}\nu \left( \dot{x}\right)
\int_{G/H}\left\| f^{x}\left( \dot{g}\right) \right\| _{E}^{2}\text{d}\mu
_{G/H}\left( \dot{g}\right)  \\
&=&\int_{\widehat{G}/H^{\perp }}\text{d}\nu \left( \dot{x}\right)
\int_{G/H}\left\| f\left( g,\dot{x}\right) \right\| _{E}^{2}\text{d}\mu
_{G/H}\left( \dot{g}\right)  \\
&=&\int_{G/H\times \widehat{G}/H^{\perp }}\left\| f(g,\dot{x})\right\|
_{E}^{2}\text{d}\left( \mu _{G/H}\otimes \nu \right) \left( \dot{g},\dot{x} 
\right)  \\
&=&\Vert f\Vert _{\mathcal{H}^{\nu }}^{2}\text{.}
\end{eqnarray*}
By density, $\Sigma $ extends to an isometry from $\mathcal{H}^{\nu }$ to $ 
L^{2}\left( \widehat{G},\widetilde{\nu };E\right) $. Clearly, Eq.~(\ref
{Formula di sigma}) holds and it defines uniquely $\Sigma $.

The second step is computing the adjoint of $\Sigma $. Let $\varphi \in
C_{c}\left( \widehat{G};E\right) $, by standard arguments the right hand side
of
 Eq.~(\ref{star}) is a continuous function of $\left( g,\dot{x}\right) $.
Moreover, it satisfies Eq.~(\ref{inv}). We have 
\begin{equation*}
\int_{H^{\perp }}\overline{\left\langle xy,g\right\rangle }\varphi \left(
xy\right) \text{\textnormal{d}}\mu _{H^{\perp }}\left( y\right) =\overline{ 
\left\langle x,g\right\rangle }\overline{\mathcal{F}}_{G/H}^{\ast }\left(
\varphi _{x}\right) \left( \dot{g}\right) \qquad \left( g,\dot{x}\right) \in
G\times \widehat{H}\text{.}
\end{equation*}
First of all, we show that the above function is in $\mathcal{H}^{\nu }$.
Indeed, {\setlength\arraycolsep{0pt} 
\begin{eqnarray}
&&\int_{\widehat{G}/H^{\perp }}\text{d}\nu \left( \dot{x}\right)
\int_{G/H}\left\| \overline{\left\langle x,g\right\rangle }\overline{ 
\mathcal{F}}_{G/H}^{\ast }\left( \varphi _{x}\right) \left( \dot{g}\right)
\right\| _{E}^{2}\text{d}\mu _{G/H}\left( \dot{g}\right)   \notag \\
&&\quad \quad \quad =\int_{\widehat{G}/H^{\perp }}\text{d}\nu \left( \dot{x} 
\right) \int_{G/H}\left\| \overline{\mathcal{F}}_{G/H}^{\ast }\left( \varphi
_{x}\right) \left( \dot{g}\right) \right\| _{E}^{2}\text{d}\mu _{G/H}\left( 
\dot{g}\right)   \notag \\
&&\quad \quad \quad \quad (\text{unitarity~of~}\overline{\mathcal{F}}_{G/H})
\notag \\
&&\quad \quad \quad =\int_{\widehat{G}/H^{\perp }}\text{d}\nu \left( \dot{x} 
\right) \int_{H^{\bot }}\left\| \varphi _{x}\left( y\right) \right\| _{E}^{2} 
\text{d}\mu _{H^{\bot }}\left( y\right)   \notag \\
&&\quad \quad \quad =\int_{\widehat{G}/H^{\perp }}\text{d}\nu \left( \dot{x} 
\right) \int_{H^{\bot }}\left\| \varphi \left( xy\right) \right\| _{E}^{2} 
\text{d}\mu _{H^{\bot }}\left( y\right)   \notag \\
&&\quad \quad \quad =\left\| \varphi \right\| _{L^{2}\left( \widehat{G}, 
\widetilde{\nu };E\right) }^{2}\text{.}  \label{modulo quadro}
\end{eqnarray}
}Moreover, for all $f\in \mathcal{H}_{0}^{\nu }$, we have {\setlength\arraycolsep{0pt} 
\begin{eqnarray*}
\left\langle \Sigma ^{\ast }\varphi ,f\right\rangle _{\mathcal{H}^{\nu }}
&=&\left\langle \varphi ,\Sigma f\right\rangle _{L^{2}\left( \widehat{G}, 
\widetilde{\nu };E\right) } \\
&=&\int_{\widehat{G}/H^{\perp }}\text{d}\nu \left( \dot{x}\right)
\int_{H^{\perp }}\left\langle \varphi \left( xy\right) ,\left( \Sigma
f\right) \left( xy\right) \right\rangle _{E}\text{d}\mu _{H^{\perp }}\left(
y\right)  \\
&=&\int_{\widehat{G}/H^{\perp }}\text{d}\nu \left( \dot{x}\right)
\int_{H^{\perp }}\left\langle \varphi _{x}\left( y\right) ,\overline{ 
\mathcal{F}}_{G/H}\left( f^{x}\right) \left( y\right) \right\rangle _{E} 
\text{ d}\mu _{H^{\perp }}\left( y\right)  \\
&&\text{(unitarity of }\overline{\mathcal{F}}_{G/H}\text{) } \\
&=&\int_{\widehat{G}/H^{\perp }}\text{d}\nu \left( \dot{x}\right)
\int_{G/H}\left\langle \overline{\mathcal{F}}_{G/H}^{\ast }\left( \varphi
_{x}\right) \left( \dot{g}\right) ,f^{x}\left( \dot{g}\right) \right\rangle
_{E}\text{d}\mu _{G/H}\left( \dot{g}\right)  \\
\quad \quad \quad \quad  &=&\int_{\widehat{G}/H^{\perp }}\text{d}\nu \left( 
\dot{x}\right) \int_{G/H}\left\langle \overline{\mathcal{F}}_{G/H}^{\ast
}\left( \varphi _{x}\right) \left( \dot{g}\right) ,\left\langle
x,g\right\rangle f\left( g,\dot{x}\right) \right\rangle _{E}\text{d}\mu
_{G/H}\left( \dot{g}\right)  \\
&=&\int_{\widehat{G}/H^{\perp }}\text{d}\nu \left( \dot{x}\right)
\int_{G/H}\left\langle \overline{\left\langle x,g\right\rangle }\overline{ 
\mathcal{F}}_{G/H}^{\ast }\left( \varphi _{x}\right) \left( \dot{g}\right)
,f\left( g,\dot{x}\right) \right\rangle _{E}\text{d}\mu _{G/H}\left( \dot{g} 
\right)  \\
&=&\int_{G/H\times \widehat{G}/H^{\perp }}\left\langle \overline{ 
\left\langle x,g\right\rangle }\overline{\mathcal{F}}_{G/H}^{\ast }\left(
\varphi _{x}\right) \left( \dot{g}\right) ,f\left( g,\dot{x}\right)
\right\rangle _{E}\text{d}\left( \mu _{G/H}\otimes \nu \right) \left( \dot{g} 
,\dot{x}\right) \text{.}
\end{eqnarray*}
}Since $\mathcal{H}_{0}^{\nu }$ is dense, Eq.~(\ref{star}) follows. By Eq.~(\ref{modulo quadro}) $\Sigma ^{\ast }$ is isometric, hence $\Sigma $ is
unitary.

Finally, we show the intertwining property. Let $a\in G$ and $f\in \mathcal{H 
}_{0}^{\nu }$. Then $\lambda ^{\nu }\left( a\right) f\in \mathcal{H} 
_{0}^{\nu }$, and so one has 
\begin{eqnarray*}
\left( \Sigma \lambda ^{\nu }\left( a\right) f\right) \left( x\right) 
&=&\int_{G/H}\left\langle x,g\right\rangle f\left( a^{-1}g,\dot{x}\right) 
\text{d}\mu _{G/H}\left( \dot{g}\right)  \\
&=&\left\langle x,a\right\rangle \int_{G/H}f^{x}\left( a^{-1}[\dot{g} 
]\right) \text{d}\mu _{G/H}\left( \dot{g}\right)  \\
&&(\dot{g}\rightarrow a\left[ \dot{g}\right] ) \\
&=&\left\langle x,a\right\rangle \int_{G/H}\left\langle x,g\right\rangle
f\left( g,\dot{x}\right) \text{d}\mu _{G/H}\left( \dot{g}\right)  \\
&=&\left( \Lambda ^{\nu }\left( a\right) \Sigma f\right) \left( x\right)
\qquad x\in \widehat{G}\text{.}
\end{eqnarray*}
By density of $\mathcal{H}_{0}^{\nu }$, it follows that $\Sigma \lambda
^{\nu }\left( a\right) =\Lambda ^{\nu }\left( a\right) \Sigma $.
\end{proof}

Given $\omega \in C_{c}\left( G/H\right) $, let $\widetilde{M_{0}^{\nu }} 
\left( \omega \right) =\Sigma M_{0}^{\nu }\left( \omega \right) \Sigma
^{\ast }$. Then

\begin{proposition}
For all $\omega \in C_{c}\left( G/H\right) $ and $\phi \in L^{2}\left( 
\widehat{G},\widetilde{\nu };E\right) $, 
\begin{equation}
\left( \widetilde{M_{0}^{\nu }}\left( \omega \right) \phi \right) \left(
x\right) =\int_{H^{\perp }}\overline{\mathcal{F}}_{G/H}\left( \omega \right)
\left( y\right) \phi \left( xy^{-1}\right) \text{d}\mu _{H^{\perp }}\left(
y\right) \qquad x\in \widehat{G}\text{.}  \label{La PVM !}
\end{equation}
\end{proposition}

\begin{proof}
Let $\omega \in C_{c}\left( G/H\right) $. We compute the action of $
\widetilde{M_{0}^{\nu }}\left( \omega \right) $ on $C_{c}\left( \widehat{G} 
;E\right) $. If $\varphi \in C_{c}\left( \widehat{G};E\right) $, let 
\begin{equation*}
\xi \left( x\right) :=\int_{H^{\perp }}\overline{\mathcal{F}}_{G/H}\left(
\omega \right) \left( y\right) \varphi \left( xy^{-1}\right) \text{d}\mu
_{H^{\perp }}\left( y\right) \qquad \forall x\in \widehat{G}\text{,}
\end{equation*}
which is well defined and continuous. Moreover, for all $x\in \widehat{G}$
and $y\in H^{\perp }$, 
\begin{eqnarray}
\xi \left( xy\right)  &=&\int_{H^{\perp }}\overline{\mathcal{F}}_{G/H}\left(
\omega \right) \left( y^{\prime }\right) \varphi \left( xyy^{\prime
-1}\right) \text{d}\mu _{H^{\perp }}\left( y^{\prime }\right)   \notag \\
&=&\int_{H^{\perp }}\overline{\mathcal{F}}_{G/H}\left( \omega \right) \left(
y^{\prime }\right) \varphi _{x}\left( yy^{\prime -1}\right) \text{d}\mu
_{H^{\perp }}\left( y^{\prime }\right)   \notag \\
&=&\left( \overline{\mathcal{F}}_{G/H}\left( \omega \right) \ast \varphi
_{x}\right) \left( y\right) \text{.}  \label{Eq. 1 in POVM}
\end{eqnarray}
Here and in the following, convolutions are always taken in $H^{\perp }$. If 
$\varphi ,\psi \in C_{c}\left( \widehat{G};E\right) $,{\setlength
\arraycolsep{0pt} 
\begin{eqnarray*}
&&\left\langle \widetilde{M_{0}^{\nu }}\left( \omega \right) \varphi ,\psi
\right\rangle _{L^{2}\left( \widehat{G},\widetilde{\nu };E\right)
}=\left\langle M_{0}^{\nu }\left( \omega \right) \Sigma ^{\ast }\varphi
,\Sigma ^{\ast }\psi \right\rangle _{\mathcal{H}^{\nu }} \\
&&\quad \quad =\int_{\widehat{G}/H^{\perp }}\text{d}\nu \left( \dot{x} 
\right) \int_{G/H}\text{d}\mu _{G/H}\left( \dot{g}\right) \Big\langle\omega
\left( \dot{g}\right) \overline{\left\langle x,g\right\rangle }\overline{ 
\mathcal{F}}_{G/H}^{\ast }\left( \varphi _{x}\right) \left( \dot{g}\right) , 
\overline{\left\langle x,g\right\rangle }\times  \\
&&\quad \quad \quad \times \overline{\mathcal{F}}_{G/H}^{\ast }\left( \psi
_{x}\right) \left( \dot{g}\right) \Big\rangle_{E} \\
&&\quad \quad =\int_{\widehat{G}/H^{\perp }}\text{d}\nu \left( \dot{x} 
\right) \int_{G/H}\text{d}\mu _{G/H}\left( \dot{g}\right) \left\langle
\omega \left( \dot{g}\right) \overline{\mathcal{F}}_{G/H}^{\ast }\left(
\varphi _{x}\right) \left( \dot{g}\right) ,\overline{\mathcal{F}} 
_{G/H}^{\ast }\left( \psi _{x}\right) \left( \dot{g}\right) \right\rangle
_{E} \\
&&\quad \quad \quad \text{(unitarity of }\overline{\mathcal{F}}_{G/H}\text{
and properties of convolution)} \\
&&\quad \quad =\int_{\widehat{G}/H^{\perp }}\text{d}\nu \left( \dot{x} 
\right) \int_{H^{\perp }}\text{d}\mu _{H^{\perp }}\left( y\right)
\left\langle \left( \overline{\mathcal{F}}_{G/H}\left( \omega \right) \ast
\varphi _{x}\right) \left( y\right) ,\psi _{x}\left( y\right) \right\rangle
_{E} \\
&&\quad \quad =\int_{\widehat{G}/H^{\perp }}\text{d}\nu \left( \dot{x} 
\right) \int_{H^{\perp }}\text{d}\mu _{H^{\perp }}\left( y\right)
\left\langle \xi \left( xy\right) ,\psi \left( xy\right) \right\rangle _{E} 
\text{,}
\end{eqnarray*}
} hence Eq.~(\ref{La PVM !}) holds on $C_{c}\left( \widehat{G};E\right) $.

Let now $\phi \in L^{2}\left( \widehat{G},\widetilde{\nu };E\right) $. Since 
\begin{equation*}
\left\| \phi \right\| _{L^{2}\left( \widehat{G},\widetilde{\nu };E\right)
}^{2}=\int_{\widehat{G}/H^{\perp }}\text{d}\nu \left( \dot{x}\right)
\int_{H^{\perp }}\Vert \phi (xy)\Vert _{E}^{2}\text{d}\mu _{H^{\perp
}}\left( y\right) <+\infty \text{,}
\end{equation*}
by virtue of Fubini theorem there is a $\nu $-negligible set $Y_{1}\subset 
\widehat{G}/H^{\perp }$ such that, for all $x\in \widehat{G}$ with $\dot{x} 
\not\in Y_{1}$, $\phi _{x}\in L^{2}\left( H^{\perp },\mu _{H^{\perp
}};E\right) $. Moreover, using the definition of $\widetilde{\nu }$, one can
check that $\pi ^{-1}\left( Y_{1}\right) $ is $\widetilde{\nu }$-negligible.
Then, for $\widetilde{\nu }$-almost all $x\in \widehat{G}$, $\phi _{x}$ is
in $L^{2}\left( H^{\perp },\mu _{H^{\perp }};E\right) $. We observe that the
map 
\begin{equation*}
\dot{g}\longmapsto \omega \left( \dot{g}\right) \left( \overline{\mathcal{F}} 
_{G/H}^{\ast }\left( \phi _{x}\right) \right) \left( \dot{g}\right) 
\end{equation*}
is then in $\left( L^{1}\cap L^{2}\right) \left( G/H,\mu _{G/H};E\right) $
for $\widetilde{\nu }$-almost all $x\in \widehat{G}$, hence its Fourier
cotransform is continuous, and we have 
\begin{eqnarray}
\overline{\mathcal{F}}_{G/H}\left( \omega \overline{\mathcal{F}}_{G/H}^{\ast
}\left( \phi _{x}\right) \right) \left( e\right)  &=&\left( \overline{ 
\mathcal{F}}_{G/H}\left( \omega \right) \ast \phi _{x}\right) \left(
e\right)   \notag \\
&=&\int_{H^{\perp }}\overline{\mathcal{F}}_{G/H}\left( \omega \right) \left(
y\right) \phi \left( xy^{-1}\right) \text{d}\mu _{H^{\perp }}\left( y\right) 
\text{.}  \label{Eq. 2 in POVM}
\end{eqnarray}
Now, we let $\left( \varphi _{k}\right) _{k\geq 1}$ be a sequence in $ 
C_{c}\left( \widehat{G};E\right) $ converging to $\phi $ in $L^{2}\left( 
\widehat{G},\widetilde{\nu };E\right) $. Then
\begin{equation*}
\int_{\widehat{G}/H^{\perp }}\text{d}\nu \left( \dot{x}\right)
\int_{H^{\perp }}\left\| \left( \varphi _{k}\right) _{x}\left( y\right)
-\phi _{x}\left( y\right) \right\| _{E}^{2}\text{d}\mu _{H^{\perp }}\left(
y\right) \longrightarrow 0
\end{equation*}
and so, possibly passing to a subsequence, there is a $\nu $-negligible set $ 
Y_{2}\subset \widehat{G}/H^{\perp }$ such that
\begin{equation*}
\int_{H^{\perp }}\left\| \left( \varphi _{k}\right) _{x}\left( y\right)
-\phi _{x}\left( y\right) \right\| _{E}^{2}\text{d}\mu _{H^{\perp }}\left(
y\right) \longrightarrow 0
\end{equation*}
for all $x\in \widehat{G}$ with $\dot{x}\not\in Y_{2}$. 
This fact means that, for $\widetilde{\nu }$-almost all $x\in \widehat{G}$, 
\begin{equation*}
\left( \varphi _{k}\right) _{x}\longrightarrow \phi _{x}
\end{equation*}
in $L^{2}\left( H^{\perp },\mu _{H^{\perp }};E\right) $. It follows that 
\begin{equation*}
\omega \overline{\mathcal{F}}_{G/H}^{\ast }\left( \left( \varphi _{k}\right)
_{x}\right) \longrightarrow \omega \overline{\mathcal{F}}_{G/H}^{\ast
}\left( \phi _{x}\right) 
\end{equation*}
in $L^{1}\left( G/H,\mu _{G/H};E\right) $. Then, for $\widetilde{\nu }$-almost all $x\in \widehat{G}$, 
\begin{equation*}
\overline{\mathcal{F}}_{G/H}\left( \omega \overline{\mathcal{F}}_{G/H}^{\ast
}\left( \left( \varphi _{k}\right) _{x}\right) \right) \longrightarrow 
\overline{\mathcal{F}}_{G/H}\left( \omega \overline{\mathcal{F}}_{G/H}^{\ast
}\left( \phi _{x}\right) \right) 
\end{equation*}
uniformly, and, using Eqs.~(\ref{Eq. 1 in POVM}), (\ref{Eq. 2 in POVM}),
{\setlength\arraycolsep{0pt}
\begin{eqnarray*}
&&\left( \widetilde{M_{0}^{\nu }}\left( \omega \right) \varphi _{k}\right)
\left( x\right) =\overline{\mathcal{F}}_{G/H}\left( \omega \overline{ 
\mathcal{F}}_{G/H}^{\ast }\left( \left( \varphi _{k}\right) _{x}\right)
\right) \left( e\right) \longrightarrow  \\
&&\quad \quad \quad \longrightarrow \overline{\mathcal{F}}_{G/H}\left(
\omega \overline{\mathcal{F}}_{G/H}^{\ast }\left( \phi _{x}\right) \right)
\left( e\right) =\int_{H^{\perp }}\overline{\mathcal{F}}_{G/H}\left( \omega
\right) \left( y\right) \phi \left( xy^{-1}\right) \text{d}\mu _{H^{\perp
}}\left( y\right) \text{.}
\end{eqnarray*}
}Since $\widetilde{M_{0}^{\nu }}\left( \omega \right) \varphi _{k}$ converges to $ 
\widetilde{M_{0}^{\nu }}\left( \omega \right) \phi $ in $L^{2}\left( 
\widehat{G},\widetilde{\nu };E\right) $, Eq.~(\ref{La PVM !}) follows
from unicity of the limit.
\end{proof}

\section{Characterisation of covariant POVMs\label{Sec. 3}}
We fix in the following an \emph{infinite dimensional} Hilbert space $E$. 
According to the results of the previous sections, the generalised
imprimitivity theorem for Abelian groups can be stated in the following way.

\begin{theorem}
\label{GIT} A POVM $M$ based on $G/H$ and acting on $\mathcal{H}$ is
covariant with respect to $U$ if and only if there exist a measure $\nu $ on 
$\widehat{G}/H^{\perp }$ and an isometry $W$ intertwining $U$ with $\Lambda
^{\nu }$ such that 
\begin{equation*}
M\left( \omega \right) =W^{\ast }\widetilde{M_{0}^{\nu }}\left( \omega
\right) W
\end{equation*}
for all $\omega \in C_{c}\left( G/H\right) $.
\end{theorem}

To get an explicit form of $W$, we assume that $U$ acts diagonally on $\mathcal{H}$. This means that
$\mathcal{H}$ is the orthogonal sum of invariant subspaces
\begin{equation}
\mathcal{H}=\tbigoplus_{k\in I}L^{2}\left( \widehat{G},\rho
_{k};F_{k}\right) \text{,}  \label{decomp. di U,H 2}
\end{equation}
where $I$ is a denumerable set, $\left( \rho _{k}\right) _{k\in I}$ is a
family of measures on $\widehat{G}$, $\left( F_{k}\right) _{k\in I}$ is a
family of Hilbert spaces, and the action of $U$ is given by 
\begin{equation*}
\left( U\left( g\right) \phi _{k}\right) \left( x\right) =\left\langle
x,g\right\rangle \phi _{k}\left( x\right) \qquad x\in \widehat{G} \text{,}
\end{equation*}
where $\phi _{k}\in L^{2}\left( \widehat{G} 
,\rho _{k};F_{k}\right) $ and $g\in G$. We will denote by $P_{k}$ the orthogonal projector onto
the subspace $L^{2}\left( \widehat{G},\rho _{k};F_{k}\right) $.

The assumption~(\ref{decomp. di U,H 2}) is not restrictive. Indeed, it is
well known  that there are a family of disjoint measures $ 
(\rho_k)_{k\in \mathbb{N}\cup\{\infty\}}$ and a family of Hilbert spaces
$(F_k)_{k\in \mathbb{N}\cup\{\infty\}}$ such
 that $\dim F_{k}=k$ and, up to a
unitary equivalence, Eq.~(\ref{decomp. di
 U,H 2}) holds.

Given the decomposition~(\ref{decomp. di U,H 2}), let $\rho$ be a measure on 
$\widehat{G}$ such that 
\begin{equation}  \label{implic.}
\rho(N)=0 \Longleftrightarrow \rho_k(N)=0 \ \ \forall k\in I.
\end{equation}
We recall that the equivalence class of $\rho$ is uniquely defined by the
family $(\rho_k)_{k\in I}$.

Finally, we observe also that the equivalence class of $\rho $ is
independent of the choice of decomposition (\ref{decomp. di U,H 2}). Indeed,
if $G$ acts diagonally on another decomposition 
\begin{equation*}
\mathcal{H}=\tbigoplus_{k\in I^{\prime}}L^{2}\left( \widehat{G},{ 
\rho^{\prime}}_{k};{F^{\prime}}_{k}\right) \text{,}
\end{equation*}
then 
\begin{equation*}
\rho _{k}^{\prime }\left( N\right) =0\text{ }\forall k\in I^{\prime
}\Longleftrightarrow \rho _{k}\left( N\right) =0\text{ }\forall k\in I\text{.}
\end{equation*}

It follows that the representation $U$ defines uniquely an equivalence class 
$\mathcal{C}_U$ of measures $\rho $ such that relation~(\ref{implic.})
holds. Chosen in this equivalence class a \emph{finite} measure $\rho$, we
denote by $\mathcal{C}_U^{\pi}$ the equivalence class of the image measure $ 
\rho^{\pi}$. Clearly $\mathcal{C}_U^\pi$ depends only on $\mathcal{C}_U$.

We now give the central result of this section.

\begin{theorem}
\label{Prop. centr.} Let $U$ be a representation of $G$ acting diagonally on 
\begin{equation*}
\mathcal{H}=\tbigoplus_{k\in I}L^{2}\left( \widehat{G},\rho
_{k};F_{k}\right) .
\end{equation*}
Given $\nu _{U}\in \mathcal{C}_{U}^{\pi }$, let $\widetilde{\nu }_{U}$ be
the measure given by Eq.~$(\ref{def. nu tilde})$. The representation $U$
admits covariant positive operator valued measures based on $G/H$ if and
only if, for all $k\in I$, $\rho _{k}$ has density with respect to $ 
\widetilde{\nu }_{U}$. In this case, for every $k\in I$, let $\alpha _{k}$
be the densities of $\rho _{k}$ with respect to $\widetilde{\nu }_{U}$.

Let $E$ be a fixed infinite dimensional Hilbert space. For each $k\in I$,
let 
\begin{equation*}
\widehat{G}\ni x\longmapsto W_{k}\left( x\right) \in \mathcal{L}\left(
F_{k};E\right) 
\end{equation*}
be a weakly measurable map such that $W_{k}\left( x\right) $ are isometries
for $\rho _{k}$-almost all $x\in \widehat{G}$. For $\omega \in C_{c}\left(
G/H\right) $, let $M\left( \omega \right) $ be the operator given by 
\begin{eqnarray}
\left( P_{j}M\left( \omega \right) P_{k}\phi \right) \left( x\right) 
&=&\int_{H^{\perp }}\text{d}\mu _{H^{\perp }}\left( y\right) \overline{ 
\mathcal{F}}_{G/H}\left( \omega \right) \left( y\right) \sqrt{\frac{\alpha
_{k}\left( xy^{-1}\right) }{\alpha _{j}\left( x\right) }}\times   \notag \\
&&\times W_{j}\left( x\right) ^{\ast }W_{k}\left( xy^{-1}\right) \left(
P_{k}\phi \right) \left( xy^{-1}\right) \qquad x\in \widehat{G},
\label{eq. di M buona}
\end{eqnarray}
for all $\phi \in \mathcal{H}$ and $k,j\in I$. Then, $M$ is a POVM covariant
with respect to $U$.

Conversely, any POVM based on $G/H$ and covariant with respect to $U$ is of
the form given by Eq.~$(\ref{eq. di M buona})$.
\end{theorem}

We add some comments before the proof of the theorem.

\begin{remark}
We observe that Eq.~$(\ref{eq. di M buona})$ is invariant with respect to
the choice of the measure $\nu _{U}\in \mathcal{C}_{U}^{\pi }$. Indeed, let $ 
\nu _{U}^{\prime }\in \mathcal{C}_{U}^{\pi }$, and $\beta >0$ be the density
of $\nu _{U}$ with respect to $\nu _{U}^{\prime }$. Clearly 
\begin{equation*}
\widetilde{\nu _{U}}=\left( \beta \circ \pi \right) \widetilde{\nu
_{U}^{\prime }}\text{,}
\end{equation*}
so that the densities $\alpha _{k}^{\prime }$ of $\rho _{k}$ with respect to 
$\widetilde{\nu _{U}^{\prime }}$ are 
\begin{equation*}
\alpha _{k}^{\prime }=\left( \beta \circ \pi \right) \alpha _{k}\text{.}
\end{equation*}
It follows that Eq.~$(\ref{eq. di M buona})$ does not depend on the choice
of $\nu _{U}\in \mathcal{C}_{U}^{\pi }$.
\end{remark}

\begin{corollary}
\label{banale} Let $H$ be the trivial subgroup $\{e\}$. The representation $U
$ admits covariant positive operator valued measures based on $G$ if and only if
the measures $\rho _{k}$ have density with respect to the Haar measure $\mu
_{\widehat{G}}$. In this case, the functions $\alpha _{k}$ in Eq.~$(\ref{eq.
di M buona})$ are the densities of $\rho _{k}$ with respect to $\mu _{ 
\widehat{G}}$.
\end{corollary}

\begin{remark}
The content of the previous corollary was first shown by Holevo in Ref.~\cite{holevo1}
for non-normalised POVM. In order to compare the two results observe that,
if $\phi \in \left( L^{1}\cap L^{2}\right) \left( \widehat{G},\rho
_{k};F_{k}\right) $ and $\psi \in \left( L^{1}\cap L^{2}\right) \left( 
\widehat{G},\rho _{j};F_{j}\right) $, Eq.~$(\ref{eq. di M buona})$ becomes {\setlength\arraycolsep{0pt} 
\begin{eqnarray*}
&&\left\langle M\left( \omega \right) \phi ,\psi \right\rangle _{\mathcal{H} 
}=\int_{G}\text{d}\mu _{G}\left( g\right) \omega \left( g\right) \int_{ 
\widehat{G}\times \widehat{G}}\left\langle x,g\right\rangle \overline{ 
\left\langle y,g\right\rangle }\sqrt{\alpha _{k}\left( y\right) \alpha
_{j}\left( x\right) }\times  \\
&&\quad \quad \quad \quad \times \left\langle W_{j}\left( x\right) ^{\ast
}W_{k}\left( y\right) \phi \left( y\right) ,\psi \left( x\right)
\right\rangle \text{d}\left( \mu _{\widehat{G}}\otimes \mu _{\widehat{G} 
}\right) \left( x,y\right)  \\
&&\quad \quad \quad =\int_{G}\text{d}\mu _{G}\left( g\right) \omega \left(
g\right) \int_{\widehat{G}\times \widehat{G}}K_{U\left( g^{-1}\right) \psi
,U\left( g^{-1}\right) \phi }\left( x,y\right) \text{d}\left( \mu _{\widehat{ 
G}}\otimes \mu _{\widehat{G}}\right) \left( x,y\right) \text{,}
\end{eqnarray*}
}where 
\begin{equation*}
K_{\psi ,\phi }\left( x,y\right) =\sqrt{\alpha _{k}\left( y\right) \alpha
_{j}\left( x\right) }\left\langle W_{k}\left( y\right) \phi \left( y\right)
,W_{j}\left( x\right) \psi \left( x\right) \right\rangle 
\end{equation*}
is a bounded positive definite measurable field of forms (compare with Eqs.~$(4.2)$ and~$(4.3)$ in Ref.~\cite{holevo1}).
\end{remark}

In order to prove Theorem~\ref{Prop. centr.}, we need the following lemma.

\begin{lemma}
\label{prop. 2.2} Let $\rho $ be a finite measure on $\widehat{G}$. Assume
that there is a measure $\nu $ on $\widehat{G}/H^{\perp }$ such that $\rho $
has density with respect to $\widetilde{\nu }$. Then $\rho $ has density
with respect to $\widetilde{\rho ^{\pi }}$. In this case, $\nu $ uniquely
decomposes as 
\begin{equation*}
\nu =\nu _{1}+\nu _{2}\text{,}
\end{equation*}
where $\nu _{1}$ is equivalent to $\rho ^{\pi }$ and $\nu _{2}\perp \rho
^{\pi }$.
\end{lemma}

\begin{proof}
Suppose that $\nu $ is a measure on $\widehat{G}/H^{\perp }$ such that $\rho
=\alpha \widetilde{\nu }$, where $\alpha $ is a non-negative $\widetilde{\nu
}$-integrable function on $\widehat{G}$. Then, for all $\varphi \in
C_{c}\left( \widehat{G}/H^{\perp }\right) $, 
 \begin{eqnarray*}
\rho ^{\pi }\left( \varphi \right)  &=&\int_{\widehat{G}}\varphi \left( \pi
\left( x\right) \right) \text{d}\rho \left( \dot{x}\right)  \\
&=&\int_{\widehat{G}/H^{\perp }}\text{d}\nu \left( \dot{x}\right)
\int_{H^{\perp }}\varphi \left( \dot{x}\right) \alpha \left( xy\right) \text{d}\mu _{H^{\perp }}\left( y\right)  \\
&=&\int_{\widehat{G}/H^{\perp }}\varphi \left( \dot{x}\right) \alpha
^{\prime }\left( \dot{x}\right) \text{d}\nu \left( \dot{x}\right) \text{,}
\end{eqnarray*}
where the function 
\begin{equation*}
\alpha ^{\prime }\left( \dot{x}\right) :=\int_{H^{\perp }}\alpha \left(
xy\right) d\mu _{H^{\perp }}\left( y\right) \geq 0
\end{equation*}
is $\nu $-integrable by virtue of Fubini theorem. It follows that 
\begin{equation}
\rho ^{\pi }=\alpha ^{\prime }\nu \text{.}  \label{Mostra}
\end{equation}
Using Lebesgue theorem, we can uniquely decompose 
\begin{equation*}
\nu =\nu _{1}+\nu _{2}\text{,}
\end{equation*}
where $\nu _{1}$ has base $\rho ^{\pi }$ and $\nu _{2}\perp \rho ^{\pi}$.
From Eq.~(\ref{Mostra}), it follows that $\nu _{1}$ and $\rho ^{\pi }$ are
equivalent, and this proves the second statement of the lemma. If $A,B\in 
\mathcal{B}\left( \widehat{G}/H^{\perp }\right) $ are disjoint sets such
 that
$\nu _{2}$ is concentrated in $A$ and $\nu _{1}$ is concentrated in 
 $B$,
then $\widetilde{\nu _{2}}$ and $\widetilde{\nu _{1}}$ are
 respectively
concentrated in the disjoint sets $\widetilde{A}=\pi
 ^{-1}\left( A\right) $
and $\widetilde{B}=\pi ^{-1}\left( B\right) $. By
 definition of $\rho ^{\pi
}$, we also have 
 \begin{equation*}
 \rho \left( \widetilde{A}\right) =\rho
^{\pi }\left( A\right) =0\text{.}
 \end{equation*}
Since $\rho $ has density with respect to $\widetilde{\nu }=\widetilde{\nu
_{1}}+\widetilde{\nu _{2}}$ and $\widetilde{\nu _{2}}$ is concentrated in $ 
\widetilde{A}$, it follows that $\rho $ has density with respect
to $\widetilde{\nu _{1} }\cong \widetilde{\rho ^{\pi }
 }$. The claim is now clear.
 \end{proof}

\begin{proof}[Proof of Theorem~$\ref{Prop. centr.}$]
Let $\rho $ be a finite measure in $\mathcal{C}_{U}$. By virtue of Theorem 
\ref{GIT}, $U$ admits a covariant POVM $\Longleftrightarrow $ there exists a
measure $\nu $ in $\widehat{G}/H^{\perp }$ such that $U$ is a
subrepresentation of $\Lambda ^{\nu }\Longleftrightarrow $ each measure $ 
\rho _{k}$ has density with respect to $\widetilde{\nu }$ $ 
\Longleftrightarrow $ $\rho $ has density with respect to $\widetilde{\nu }$. From Lemma \ref{prop. 2.2}, $U$ admits a covariant POVM if and only if $ 
\rho $ has density with respect to $\widetilde{\rho ^{\pi }}$. Since $\rho
^{\pi }\in \mathcal{C}_{U}^{\pi }$, the first claim follows.

Let now $M$ be a covariant POVM. By Theorem \ref{GIT}, there is a measure $ 
\nu $ on $\widehat{G}/H^{\perp }$ and an isometry $W$ intertwining $U$ with $ 
\Lambda ^{\nu }$ such that 
\begin{equation*}
M\left( \omega \right) =W^{\ast }\widetilde{M_{0}^{\nu }}\left( \omega
\right) W\qquad \forall \omega \in C_{c}\left( G/H\right) \text{.}
\end{equation*}
Using Lemma \ref{prop. 2.2}, we (uniquely) decompose 
\begin{equation*}
\nu =\nu _{1}+\nu _{2}\text{,}
\end{equation*}
where $\nu _{1}$ is equivalent to $\nu _{U}$ and $\nu _{2}\perp \nu _{U}$.
Then we have 
\begin{eqnarray*}
\sigma ^{\nu } &\cong &\sigma ^{\nu _{U}}\oplus \sigma ^{\nu
_{2}}\Longrightarrow  \\
&\Longrightarrow &\left( \Lambda ^{\nu },\widetilde{M_{0}^{\nu }}\right)
\cong \left( \Lambda ^{\nu _{U}},\widetilde{M_{0}^{\nu _{U}}}\right) \oplus
\left( \Lambda ^{\nu _{2}},\widetilde{M_{0}^{\nu _{2}}}\right) \text{,}
\end{eqnarray*}
i.e.~the imprimitivity system $\left( \Lambda ^{\nu },\widetilde{M_{0}^{\nu } 
}\right) $ preserves the decomposition 
\begin{equation*}
L^{2}\left( \widehat{G},\widetilde{\nu };E\right) \cong L^{2}\left( \widehat{ 
G},\widetilde{\nu _{U}};E\right) \oplus L^{2}\left( \widehat{G},\widetilde{ 
\nu _{2}};E\right) \text{.}
\end{equation*}
Moreover, since each $\rho _{k}$ has density with respect to $\widetilde{\nu
_{U}}$ and $\widetilde{\nu _{U}}$ is disjoint from $\widetilde{\nu _{2}}$,
it follows that $W\left( \mathcal{H}\right) \subset L^{2}\left( \widehat{G}, 
\widetilde{\nu _{U}};E\right) $, then we can always assume that the measure $
\nu $ on $\widehat{G}/H^{\perp }$ which occurs in Theorem \ref{GIT} is $\nu
_{U}$.

We now characherise the form of $W$. For $k\in I$, we can always fix an
isometry $T_{k}:F_{k}\longrightarrow E$ such that $T_{k}\left( F_{k}\right) $
are mutually orthogonal subspaces of $E$. Hence, if we define, for $\phi
_{k}\in L^{2}\left( \widehat{G},\rho _{k};F_{k}\right) $, 
\begin{equation*}
\left( T\phi _{k}\right) \left( x\right) :=\sqrt{\alpha _{k}\left( x\right) } 
T_{k}\phi _{k}\left( x\right) \qquad x\in \widehat{G}\text{,}
\end{equation*}
$T$ is an isometry intertwining $U$ with $\Lambda ^{\nu _{U}}$. We define $ 
W_{k}=WP_{k}$. The operator $V=WT^{\ast }$ is a partial isometry commuting
with $\Lambda ^{\nu _{U}}$, hence there exists a weakly measurable
correspondence $\widehat{G}\ni x\longmapsto V\left( x\right) \in \mathcal{L} 
\left( E\right) $ such that $V\left( x\right) $ are partial isometries for $ 
\widetilde{\nu _{U}}$-almost all $x\in \widehat{G}$ and 
\begin{equation*}
\left( V\phi \right) \left( x\right) =V\left( x\right) \phi \left( x\right)
\qquad x\in \widehat{G}\text{,}
\end{equation*}
where $\phi \in L^{2}\left( \widehat{G},\widetilde{\nu _{U}};E\right) $. We
have $W=WT^{\ast }T=VT$, then 
\begin{eqnarray}
\left( W_{k}\phi _{k}\right) \left( x\right)  &=&\sqrt{\alpha _{k}\left(
x\right) }V\left( x\right) T_{k}\phi _{k}\left( x\right)   \notag \\
&=&\sqrt{\alpha _{k}\left( x\right) }W_{k}\left( x\right) \phi _{k}\left(
x\right) \qquad x\in \widehat{G}\text{,}  \label{Eq. di W}
\end{eqnarray}
where we set 
\begin{equation*}
W_{k}\left( x\right) =V\left( x\right) T_{k}\qquad \forall x\in \widehat{G} 
\text{.}
\end{equation*}
Since $W$ is isometric, then $W_{k}^{\ast }W_{k}$ is the identity operator
on $L^{2}\left( \widehat{G},\rho _{k};F_{k}\right) $, hence 
\begin{equation*}
T_{k}^{\ast }V\left( x\right) ^{\ast }V\left( x\right) T_{k}=I_{k}\qquad
x\in \widehat{G}
\end{equation*}
$\rho _{k}$-almost everywhere, where $I_{k}$ is the identity operator on $ 
F_{k}$. Since $T_{k}$ is isometric and $V\left( x\right) $ is a partial
isometry for $\widetilde{\nu _{U}}$-almost every $x\in \widehat{G}$ (that is
for $\rho _{k}$-almost every $x\in \widehat{G}$), it follows that $V\left(
x\right) ^{\ast }V\left( x\right) $ is the identity on $\limfunc{ran}T_{k}$
and that $W_{k}\left( x\right) $ is isometric, for $\rho _{k}$-almost every $ 
x\in \widehat{G}$. Weak measurability of the maps $x\longmapsto W_{k}\left(
x\right) $ is immediate.

The explicit form of $M$ is then given by 
\begin{eqnarray*}
\left( P_{j}M\left( \omega \right) P_{k}\phi \right) \left( x\right) 
&=&\left( W_{j}^{\ast }\widetilde{M^{\nu }_{0}}\left( \omega \right)
W_{k}\phi \right) \left( x\right)  \\
&=&\frac{1}{\sqrt{\alpha _{j}\left( x\right) }}W_{j}\left( x\right) ^{\ast
}\int_{H^{\perp }}\overline{\mathcal{F}}_{G/H}\left( \omega \right) \left(
y\right) \times  \\
&&\times \sqrt{\alpha _{k}\left( xy^{-1}\right) }W_{k}\left( xy^{-1}\right)
\left( P_{k}\phi \right) \left( xy^{-1}\right) \text{d}\mu _{H^{\perp
}}\left( y\right) \quad x\in \widehat{G}\text{,}
\end{eqnarray*}
where $\phi \in \mathcal{H}$, $\omega \in C_{c}\left( G/H\right) $.

Conversely, let $\widehat{G}\ni x\longmapsto W_{k}\left( x\right) \in 
\mathcal{L}\left( F_k;E\right) $ be a weakly measurable map such that $ 
W_{k}\left( x\right) $ are isometries for $\rho _{k}$-almost every $x\in 
\widehat{G}$ and for all $k\in I$. We define, for $\phi _{k}\in L^{2}\left( 
\widehat{G},\rho _{k};F_{k}\right) $, 
\begin{equation*}
\left( W\phi _{k}\right) \left( x\right) :=\sqrt{\alpha _{k}\left( x\right) } 
W_{k}\left( x\right) \phi _{k}\left( x\right) \qquad \forall x\in \widehat{G} 
\text{,}
\end{equation*}
then $W$ is clearly an intertwining isometry between $U$ and $\Lambda ^{\nu
_{U}}$ and Eq.~(\ref{eq. di M buona}) defines a covariant POVM.
\end{proof}

We now study the problem of equivalence of covariant POVMs. To simplify the
exposition, we assume that the measures 
$\rho _{k}$ in decomposition (\ref{decomp. di U,H 2}) are orthogonal.
 
Let $M$ and $M^{\prime }$ be two covariant positive operator valued measures
that are equivalent, i.e.~there exists an unitary
operator $S:\mathcal{H}\longrightarrow \mathcal{H}$ such that 
\begin{eqnarray}
SU\left( g\right) &=&U\left( g\right) S\qquad \forall g\in G\text{,}
\label{condiz. 1} \\
SM\left( \omega\right) &=&M^{\prime }\left( \omega\right) S\qquad \forall \omega\in 
C_c(G/H) \text{.}  \label{condiz. 2}
\end{eqnarray}
We have the following result.
\begin{proposition}
Let $\left( W_{j}\right) _{j\in I}$ and $\left(
W_{j}^{\prime }\right) _{j\in I}$ be families of maps such
that Eq.~$(\ref{eq. di M buona})$ holds for $M$ and $M'$, respectively. 

The POVMs $M$ and $M^{\prime }$
 are equivalent if and only if, for each $k\in
I$, there exists a weakly
 measurable map $x\longmapsto S_{k}\left( x\right)
\in \mathcal{L}\left(
 F_{k}\right) $ such that $S_{k}\left( x\right) $ are
unitary operators for $ 
 \rho_{k}$-almost all $x$ and 
 \begin{equation}
\sqrt{\alpha _{k}\left( xy\right) }W_{j}\left( x\right) ^{\ast }W_{k}\left(
xy\right) =\sqrt{\alpha _{k}\left( xy\right) }S_{j}\left( x\right) ^{\ast
}W_{j}^{\prime }\left( x\right) ^{\ast }W_{k}^{\prime }\left( xy\right)
S_{k}\left( xy\right)   \label{Equivalenza}
\end{equation}
for $\left( \rho _{j}\otimes \mu _{H^{\perp }}\right) $-almost all $\left(
x,y\right) $.
\end{proposition}

\begin{proof}
By virtue of condition (\ref{condiz. 1}) and orthogonality of the measures $ 
\rho _{k}$, $S$ preserves decomposition (\ref{decomp. di U,H 2}). Moreover,
for each $k\in I$, there exists a weakly measurable map $x\longmapsto
S_{k}\left( x\right) \in \mathcal{L}\left( F_{k}\right) $ such that $ 
S_{k}\left( x\right) $ is unitary for $\rho _{k}$-almost all $x$ and, if $ 
\phi _{k}\in L^{2}\left( \widehat{G},\rho _{k};F_{k}\right) $, 
\begin{equation*}
\left( S\phi _{k}\right) \left( x\right) =S_{k}\left( x\right) \phi
_{k}\left( x\right) \qquad x\in \widehat{G}\text{.}
\end{equation*}
Condition (\ref{condiz. 2})\ is equivalent to 
\begin{equation*}
P_{j}M\left( \omega \right) P_{k}\phi =P_{j}S^{\ast }M^{\prime }\left(
\omega \right) SP_{k}\phi 
\end{equation*}
for all $\phi \in \mathcal{H}$, $\omega \in C_{c}\left( G/H\right) $ and $ 
j,k\in I$. It is not restrictive to assume that the densities $\alpha _{k}$
are measurable functions. Let 
\begin{equation*}
\Omega _{j,k}\left( x,x^{\prime }\right) =\sqrt{\frac{\alpha _{k}\left(
x^{\prime }\right) }{\alpha _{j}\left( x\right) }}\left( W_{j}\left(
x\right) ^{\ast }W_{k}\left( x^{\prime }\right) -S_{j}\left( x\right) ^{\ast
}W_{j}^{\prime }\left( x\right) ^{\ast }W_{k}^{\prime }\left( x^{\prime
}\right) S_{k}\left( x^{\prime }\right) \right), 
\end{equation*}
using Eq.~(\ref{eq. di M buona}), the previous condition becomes 
\begin{equation}
\int_{H^{\perp }}\mathcal{F}_{G/H}\left( \omega \right) \left( y\right)
\Omega _{j,k}\left( x,xy^{-1}\right) \left( P_{k}\phi \right) \left(
xy^{-1}\right) \text{d}\mu _{H^{\perp }}\left( y\right) =0  \label{sopra}
\end{equation}
$\rho _{j}$-almost everywhere for all $\phi \in \mathcal{H}$, $\omega \in
C_{c}\left( G/H\right) $ and $j,k\in I$.

Let $K$ be a compact set of $\widehat{G}$ and $v\in F_{k}$. In Eq.~(\ref
{sopra}) we choose 
$$\phi=\chi _{K}\, v\in L^{2}\left( \widehat{G},\rho_{k};F_{k}\right) $$ 
and $\omega \in C_{c}\left( G/H\right) $ running over a denumerable 
subset dense in $L^{2}\left( G/H,\mu _{H^{\perp }}\right) $. 
It follows that there exists a $\rho _{j}$-null set $N\subset
\widehat{G}$
 such that, for all $x\notin N$, 
\begin{equation*}
\chi _{K}\left( xy^{-1}\right) \Omega _{j,k}\left( x,xy^{-1}\right) v=0
\end{equation*}
for $\mu _{H^{\perp }}$-almost all $y\in H^{\perp }$. Since $\Omega _{j,k}$
is weakly measurable, the last equation holds in a measurable subset $ 
Y\subset \widehat{G}\times H^{\perp }$ whose complement is a $\left( \rho
_{j}\otimes \mu _{H^{\perp }}\right) $-null set. Define 
\begin{equation*}
m\left( x,y\right) =xy^{-1}\qquad \forall \left( x,y\right) \in \widehat{G} 
\times H^{\perp }\text{.}
\end{equation*}
For all $\left( x,y\right) \in Y\cap m^{-1}\left( K\right) $ we then have 
\begin{equation*}
\Omega _{j,k}\left( x,xy^{-1}\right) v=0\text{.}
\end{equation*}
Since $F_k$ is separable and $\widehat{G}$ is $\sigma$-compact, we get 
\begin{equation*}
\Omega _{j,k}\left( x,xy\right) =0
\end{equation*}
for $\left( \rho _{j}\otimes \mu _{H^{\perp }}\right) $-almost all $\left(
x,y\right) \in \widehat{G}\times H^{\perp }$, that is, 
\begin{equation*}
\sqrt{\alpha _{k}\left( xy\right) }W_{j}\left( x\right) ^{\ast }W_{k}\left(
xy\right) =\sqrt{\alpha _{k}\left( xy\right) }S_{j}\left( x\right) ^{\ast
}W_{j}^{\prime }\left( x\right) ^{\ast }W_{k}^{\prime }\left( xy\right)
S_{k}\left( xy\right) 
\end{equation*}
for $\left( \rho _{j}\otimes \mu _{H^{\perp }}\right) $-almost all $\left(
x,y\right) $.

Conversely, if condition (\ref{Equivalenza}) is satisfied for all $j,k\in I$, then clearly $M$ is equivalent to $M^{\prime }$.
\end{proof}

\section{Examples}

\subsection{Generalised covariant position observables}

Let $\mathcal{H}=L^{2}\left( \mathbb{R},\text{d}x\right) $, where d$x$ is
the Lebesgue measure on $\mathbb{R}$. We consider the representation $U$ of
the group $\mathbb{R}$ acting on $\mathcal{H}$ as 
\begin{equation*}
\left( U(a)\phi \right) \left( x\right) =e^{iax}\phi \left( x\right) \qquad
x\in \mathbb{R}
\end{equation*}
for all $a\in \mathbb{R}$. By means of Fourier transform, $U$ is clearly
equivalent to the regular representation of $\mathbb{R}$. We classify 
the POVMs  based on $\mathbb{R}$ and covariant with
respect to $U$. With the notations of the previous sections, we have 
\begin{equation*}
G=\mathbb{R}\text{,\quad }H=\left\{ 0\right\} \text{,\quad }G/H=\mathbb{R} 
\text{,\quad }\widehat{G}=H^{\perp }=\mathbb{R}\text{,\quad }\widehat{G} 
/H^{\perp }=\left\{ 0\right\} \text{.}
\end{equation*}
We choose $\mu _{G/H}=\frac{1}{2\pi }$d$x$, so that $\mu _{H^{\perp }}=$ d$x$, and $E=\mathcal{H}$.

The representation $U$ is already diagonal with multiplicity equal to $1$,
so that in the decomposition (\ref{decomp. di U,H 2}) we can set $I=\left\{
1\right\} $, $\rho _{1}=$ d$x$, $F_{1}=\mathbb{C}$. Hence, by Corollary~\ref
{banale}, $U$ admits covariant POVMs based on $\mathbb{R}$ and $\alpha_{1}=1$.

According to Theorem \ref{Prop. centr.}, any
covariant POVM $M$ is defined in terms of a weakly measurable map $ 
x\longmapsto W_{1}\left( x\right) $ such that $W_{1}\left( x\right) :\mathbb{ 
C}\longrightarrow \mathcal{H}$ is an isometry for every $x\in \mathbb{R}$.
This is equivalent to selecting a weakly measurable map $x\longmapsto
h_{x}\in \mathcal{H}$, with $\left\| h_{x}\right\| _{\mathcal{H}}=1$ $ 
\forall x\in \mathbb{R}$, such that $W_{1}\left( x\right) =h_{x}$ $\forall
x\in \mathbb{R}$. Explicitly, if $\phi \in L^{2}\left( \mathbb{R},\text{d} 
x\right) $,

\begin{eqnarray*}
\left( M\left( \omega \right) \phi \right) \left( y\right) &=&\int_{\mathbb{R 
}}\overline{\mathcal{F}}_{\mathbb{R}}\left( \omega \right) \left( x\right)
\left\langle h_{y-x},h_{y}\right\rangle \phi \left( y-x\right) \text{d}x \\
&=&\int_{\mathbb{R}}\overline{\mathcal{F}}_{\mathbb{R}}\left( \omega \right)
\left( y-x\right) \left\langle h_{x},h_{y}\right\rangle \phi \left( x\right) 
\text{d}x \\
&=&\int_{\mathbb{R}}\left(\int_{\mathbb{R}}e^{ i\left( y-x\right) z}\omega
\left( z\right) \left\langle h_{x},h_{y}\right\rangle \phi \left( x\right)\ 
\frac{\text{d}z}{2\pi}\right) \text{d}x\qquad y\in \mathbb{R}.
\end{eqnarray*}

\subsection{Generalised covariant phase observables}

We give a complete characterisation of the covariance systems 
based on the one-dimensional torus
\begin{equation*}
\mathbb{T}=\left\{ z\in \mathbb{C}\mid \left| z\right| =1\right\}= \left\{
e^{i\theta}\mid \theta\in [0,2\pi]\right\}\text{.}
\end{equation*}
We have 
\begin{gather*}
G=\mathbb{T}\text{,\quad }H=\left\{ 1\right\} \text{,\quad }G/H=\mathbb{T} 
\text{,} \\
\widehat{G}=H^{\perp }=\left\{ \left( \mathbb{T}\ni z\longmapsto z^{n}\in 
\mathbb{C}\right) \mid n\in \mathbb{Z}\right\} \cong \mathbb{Z}\text{,} \\
\widehat{G}/H^{\perp }=\left\{ 1\right\} \text{.}
\end{gather*}
We choose $\mu _{G/H}=\frac{1}{2\pi}d\theta=:\mu _{\mathbb{T}}$, so that $ 
\mu _{H^{\perp }}$ is the counting measure $\mu _{\mathbb{Z}}$ on $\mathbb{Z}
$.

Let $U$ be a representation of $\mathbb{T}$. Since $\mathbb{T}$ is compact, 
we can always assume that $U$ acts diagonally on 
\begin{equation*}
\mathcal{H}=\tbigoplus_{k\in I} F_k \text{,}
\end{equation*}
where $I\subset \mathbb{Z}$, and $F_{k}$ are Hilbert spaces such that $\dim
F_{k}$ is the multiplicity of the representation $k\in \mathbb{Z}$ in $U$.
Explicitly, 
\begin{equation*}
\left( U\left( z\right) \phi _{k}\right) =z^{k}\phi_{k}
\end{equation*}
for all $z\in \mathbb{T}$ and $\phi _{k}\in F_k$.

In order to use Eq.~(\ref{decomp. di U,H 2}), we notice that $ 
F_k=L^{2}\left( \mathbb{Z},\delta_{k};F_{k}\right)$ (where $\delta _{k}$ is the Dirac measure at $k$), 
so that $\rho_{k}=\delta _{k}$, . By Corollary 
\ref{banale}, one has that $U$ admits covariant POVMs based on $\mathbb{T}$
and that $\alpha _{k}(j)=\delta _{k,j}$ (where $\delta _{k,j}$ is the Kronecker
delta).

Choose an infinite dimensional Hilbert space $E$ and, for each $k\in I$, fix
an isometry $W_{k}$ from $F_{k}$ to $E$. The corresponding covariance system
is given by

\begin{eqnarray*}
P_{j}M\left( \omega \right) P_{k}\phi & = & \overline{\mathcal{F}}_{\mathbb{T 
}}\left( \omega \right) \left( j-k\right) W_{j}^{\ast }W_{k}P_{k}\phi \\
& = & \frac{1}{2\pi}\int_{0}^{2\pi}\omega(e^{i\theta})e^{i(j-k)\theta}\,
W_{j}^{\ast }W_{k}P_k\phi\ \text{d}\theta\text{,}
\end{eqnarray*}
where $\phi \in \mathcal{H}$ and $\omega\in \mathcal{C}(\mathbb{T})$.

If $I=\mathbb{Z}$ and $\dim F_{k}=1$ $\forall k\in \mathbb{Z}$, $U$ is the
number representation and $M$ represents the phase observable (compare with
the result obtained in Ref.~\cite{Torus}).

\subsection{Covariant phase difference observables}

Let $\mu _{\mathbb{T}}$ as in the previous section. We consider the
following representation $U$ of the direct product $G=\mathbb{T}\times 
\mathbb{T}$ acting on the space $\mathcal{H}=L^{2}\left( \mathbb{T}\times 
\mathbb{T},\mu _{\mathbb{T}}\otimes \mu _{\mathbb{T}}\right) $ as
\begin{equation*}
\left( U\left( a,b\right) f\right) \left( z_{1},z_{2}\right) =f\left(
az_{1},b^{-1}z_{2}\right) \qquad \left( z_{1},z_{2}\right) \in \mathbb{ 
T\times T}
\end{equation*}
for all $\left( a,b\right) \in \mathbb{T\times T}$.

Let $H$ be the closed subgroup 
\begin{equation*}
H=\left\{ \left( a,b\right) \in \mathbb{T}\times \mathbb{T}\mid b=a\right\}
\cong \mathbb{T}.
\end{equation*}
We classify all the POVMs based on $G/H$ and covariant with respect to $U$
(for a different approach to the same problem, see Ref.~\cite{Teiko}).

We have 
\begin{gather*}
G=\mathbb{T\times T}\text{,}\quad G/H\cong \mathbb{T}\text{,\quad }\widehat{G 
}=\widehat{\mathbb{T}}\times \widehat{\mathbb{T}}\cong \mathbb{Z}\times 
\mathbb{Z}\text{,} \\
H^{\perp }=\left\{ \left( j,k\right) \in \mathbb{Z\times Z}\mid k=-j\right\}
\cong \mathbb{Z}\text{,} \\
\widehat{G}/H^{\perp }\cong \mathbb{Z}\text{.}
\end{gather*}
We fix $\mu _{G/H}=\mu _{\mathbb{T}}$, so that $\mu _{H^{\perp }}=\mu _{ 
\mathbb{Z}}$.

We choose the following orthonormal basis $\left( e_{i,j}\right) _{i,j\in 
\mathbb{Z}}$ of $\mathcal{H}$ 
\begin{equation*}
e_{i,j}\left( z_{1},z_{2}\right) =z_{1}^{i}z_{2}^{-j}\qquad \left(
z_{1},z_{2}\right) \in \mathbb{T}\times \mathbb{T}\text{,}
\end{equation*}
so that 
\begin{equation*}
U\left( a,b\right) e_{i,j}=a^{i}b^{j}e_{i,j}\qquad \forall \left( a,b\right)
\in \mathbb{T}\times \mathbb{T}\text{.}
\end{equation*}
Let $F_{i,j}=\mathbb{C}e_{i,j}$, then $U$ acts diagonally on $F_{i,j}$ as
the character $(i,j)\in \mathbb{Z}\times \mathbb{Z}$. Then, one can choose
as decomposition~(\ref{decomp. di U,H 2}) 
\begin{equation*}
\mathcal{H}=\tbigoplus_{i,j\in \mathbb{Z}}F_{i,j}\cong \tbigoplus_{i,j\in 
\mathbb{Z}}L^{2}(\mathbb{Z}\times \mathbb{Z},\delta _{i}\otimes \delta
_{j};F_{i,j})
\end{equation*}
With the notations of Section \ref{Sec. 3}, we have $I=\mathbb{Z}\times 
\mathbb{Z}$ and $\rho _{i,j}=\delta _{i}\otimes \delta _{j}$. It follows
that $\mathcal{C}_U^\pi$ is the equivalence class of $\mu _{\mathbb{Z}}$.
With the choice $\nu_U= \mu _{\mathbb{Z}}$, it follows that $\widetilde{\nu} 
=\mu _{\mathbb{Z}}\otimes \mu _{\mathbb{Z}}$. According to Theorem~(\ref
{Prop. centr.}), $U$ admits covariant POVMs and $\alpha_{i,j}(n,m)= 
\delta_{n,i}\delta_{m,j}$.

With the choice $E=\mathcal{H}$, we select a map $\left(i,j\right)
\longmapsto W_{i,j} $, where $W_{i,j}$ is an isometry from $F_{i,j}$ to $ 
\mathcal{H}$. Since $F_{i,j}$ are one dimensional, there exists a family
of vectors $\left( h_{i,j}\right) _{i,j\in \mathbb{Z}}$ in $\mathcal{H}$,
with $\left\| h_{i,j}\right\| _{\mathcal{H}}=1$ $\forall \left( i,j\right)
\in \mathbb{Z\times Z}$, such that 
\begin{equation*}
W_{i,j}e_{i,j}=h_{i,j}\qquad \forall \left( i,j\right) \in \mathbb{Z\times Z}
\text{.}
\end{equation*}
The corresponding covariant POVM $M$ is given, for every $\phi \in \mathcal{H 
}$, by 
\begin{eqnarray*}
P_{l,m} M \left( \omega \right) P_{i,j} \phi & = & \sum_{h\in \mathbb{Z}} 
\mathcal{F}_{\mathbb{T}}\left( \omega \right) \left( h\right) \delta_{l-h,i}
\delta_{m+h,j} \left\langle h_{i,j},h_{l,m}\right\rangle\
\left\langle\phi,e_{i,j}\right\rangle e_{l,m} \\
& = & \delta_{l+m,i+j}\ \mathcal{F}_{\mathbb{T}}\left( \omega \right) \left(
j-m\right) \left\langle h_{i,j},h_{l,m}\right\rangle\ \left\langle\phi,
e_{i,j}\right\rangle e_{l,m} \text{.}
\end{eqnarray*}
In particular, if $l+m=i+j$, we have 
\begin{eqnarray*}
\left\langle M\left( \omega \right) e_{i,j},e_{l,m}\right\rangle & = & 
\overline{\mathcal{F}}_{\mathbb{T}}(\omega)(j-m) \left\langle
h_{i,j},h_{l,m}\right\rangle \\
& = & \frac{1}{2\pi}\int_0^{2 \pi}\omega(e^{i\theta})e^{i(j-m)\theta}
\left\langle h_{i,j},h_{l,m}\right\rangle \text{d}\theta.
\end{eqnarray*}
If $l+m\neq i+j$, one has 
\begin{equation*}
\left\langle M\left( \omega \right) e_{i,j},e_{l,m}\right\rangle =0 \text{.}
\end{equation*}


\begin{thebibliography}{9}
\bibitem{Torus}  G.~Cassinelli, E.~De~Vito, P.~Lahti and
J.~P.~Pellonp\"{a}\"{a}, \emph{Covariant localisations in the torus and the
phase observables}, J. Math. Phys. \textbf{43}, 693-704 (2002).

\bibitem{Cass} G.~Cassinelli, E.~De~Vito, \emph{Square-integrability modulo a
subgroup}, to appear on Trans. AMS (2003). 

\bibitem{Catt}  U.~Cattaneo, \emph{On Mackey's imprimitivity
theorem},
 Comment. Math. Helvetici \textbf{54}, 629-641 (1979).

\bibitem{Dieu2}  J.~Dieudonn\'{e}, \emph{Elements d'analyse}, Vol.~\textbf{6}, Gauthiers-Villars, Paris, 1975.

\bibitem{Foll}  G.~B.~Folland, \emph{A Course in Abstract Armonic Analysis},
CRC Press, Boca Raton, 1995.

\bibitem{Teiko}  T.~Heinonen, P.~Lahti and J.~P.~Pellonp\"{a}\"{a}, \emph{Covariant phase difference observables}, to appear on J. Math. Phys. (2003).

\bibitem{holevo1}  A.~S.~Holevo, \emph{Generalized imprimitivity systems for
Abelian groups}, Izvestiya VUZ. Matematika \textbf{27}, No.~2, 53-80 (1983).

\bibitem{holevo2}  A.~S.~Holevo, \emph{On a generalization of canonical
quantization}, Math. USSR Izvestiya \textbf{28}, No.~1, 175-188 (1987).
\end{thebibliography}
\end{document}